# Narrow Gauge and Analytical Branching Strategies for Mixed Integer Programming

Fred Glover
Leeds School of Business
University of Colorado
Boulder, CO 80309-0419, USA
fred.glover@colorado.edu

Vladimir P. Shylo
V. M. Glushkov Institute of Cybernetics
National Academy of Sciences of Ukraine
40 Glushkova Ave, Kiev, Ukraine, 03187
v.shylo@gmail.com

Oleg Shylo
523 John D. Tickle Building
851 Neyland Drive
Knoxville, TN 37996-2315, USA
oshylo@utk.edu

January, 2016

## Abstract

State-of-the-art branch and bound algorithms for mixed integer programming make use of special methods for making branching decisions. Strategies that have gained prominence include modern variants of so-called strong branching (Applegate, et al.,1995) and reliability branching (Achterberg, Koch and Martin, 2005; Hendel, 2015), which select variables for branching by solving associated linear programs and exploit pseudo-costs (Benichou et al., 1971). We suggest new branching criteria and propose alternative branching approaches called *narrow gauge* and *analytical branching*. The perspective underlying our approaches is to focus on prioritization of child nodes to examine fewer candidate variables at the current node of the B&B tree, balanced with procedures to extrapolate the implications of choosing these candidates by generating a small-depth look-ahead tree. Our procedures can also be used in rules to select among *open* tree nodes (those whose child nodes have not yet been generated). We incorporate pre- and post-winnowing procedures to progressively isolate preferred branching candidates, and employ derivative (created) variables whose branches are able to explore the solution space more deeply.

## 1. Introduction

We write the mixed integer programming problem in the form

(MIP)  Minimize  $x_o = cx + dy$
       subject to  $(x,y) \in Z$ and $x \in X$

where

$Z = \{(x,y): Ax + Dy \geq b, U \geq x \geq 0\}$
$X = \{x: U \geq x \geq 0$ and $x$ integer$\}$

The vector $U$ is the n-vector of (possibly infinite) upper bounds for components of $x$, denoted by $x_j, j \in N = \{1, \ldots, n\}$. The linear programming relaxation of (MIP), which we denote by (LP), results by removing the condition $x \in X$. A solution $(x,y)$ of (LP) is called *MIP feasible* if it satisfies all the constraints of (MIP) and called *LP feasible* if it satisfies just the constraints of (LP). In case the LP relaxation has no feasible solution we say by convention that its solution is *infeasible* and specify that $x_o = $ infinity.

We let $(x^*, y^*)$ denote the best MIP feasible solution currently known, and include the requirement $x_o < x_o^*$ (e.g., $x_o \leq x_o^* - \varepsilon$) as a condition of LP feasibility, though this condition need not be explicitly added as a constraint to (LP).

Branch and bound (B&B) methods for (MIP) use branching strategies that successively impose integer lower and upper bounds $L_j^o$ and $U_j^o$ on components of $x$ to yield a series of linear programs of the form

(LP$^o$)  Minimize  $x_o = cx + dy$
          subject to  $(x,y) \in Z$ and $x \in X^o = \{x: U^o \geq x \geq L^o\}$

Each problem (LP$^o$) (and its optimal LP solution $(x^o, y^o)$) corresponds to a node of the B&B tree, with (LP) (the original relaxation of (MIP)) constituting the origin (root) node of the tree. The problem (LP$^o$) for a given tree node that is not MIP feasible gives rise to two child nodes by selecting a *fractional* variable $x_k$ (i.e., one for which the value $x_k^o$ is non-integer) in this solution and either replacing $U_k^o$ by $U_k^o := \lfloor x_k^o \rfloor$ ("rounding down" to yield a *down branch*) or replacing $L_k^o$ by $L_k^o := \lceil x_k^o \rceil$ ("rounding up" to yield an *up branch*). (The reversal of a branch, as during backtracking, occurs for a down branch by setting $L_k^o := U_k^o + 1$ and reinstating the antecedent $U_k^o$, and for an up branch by setting $U_k^o := L_k^o - 1$ and reinstating the antecedent $L_k^o$.) When both children of a parent node (LP$^o$) are infeasible, then by default the parent is infeasible too. When only one child is LP feasible, we say the associated branch is a *compulsory* branch and such a branch is automatically executed when discovered, and embodied within the current (LP$^o$) as implied restriction.

In general, (LP⁰) is often modified by making use of cutting planes or tightened bounds on variables implied by the MIP requirement that *x* is integer. In some cases, as where compulsory ranches are discovered, the tests imply bounds that are not satisfied by the current LP solution, requiring that (LP⁰) must be re-optimized. Such modifications are inherited by all descendants of (LP⁰).

Our focus in this paper is on branching strategies for identifying the branching variable $x_k$ and its preferred branching direction to grow the B&B tree. State-of-the-art branch and bound methods for solving (MIP) often make use of variants of so-called strong branching (Applegate, et al.,1995), including an extension called reliability branching (Achterberg, Koch and Martin, 2005; Hendel, 2015). The goal of these strategies is to yield effective tradeoffs between computational effort and the number of nodes generated in the B&B tree, and customarily make use of pseudo-costs (Benichou et al., 1971) to incorporate information about objective function changes produced from previous branching decisions.

We propose new branching procedures called *narrow gauge branching* and *analytical branching* that can be used to augment or replace these popular strategies. We begin by reviewing customary branching strategies in Section 2, and propose new branching choice rules as alternatives to those currently favored in the literature. In Section 3 we introduce the key ideas underlying narrow gauge branching. Section 4 addresses specific instances of narrow gauge branching that employ strategies for reducing the number of branches examined, while Section 5 introduces strategies for branching more deeply using derivative variables. The foundations of analytical branching are discussed in Section 6, which provide new approaches for generating and implementing pseudo-costs both within narrow gauge branching and more customary branching procedures. Section 7 addresses ways to exploit global relevance in branching strategies based on the notion of persistent attractiveness and taking advantage of reference sets. Finally, Section 8 presents our conclusions.

## 2. Branching Strategies – Commonly Employed and New

Let F denote the set of fractional-valued variables $x_j$ in the solution to (LP⁰)

$$F = \{j \in N: x_j^o \text{ is non-integer}\}.$$

(For convenience, we interchangeably refer to variables and their indexes as belonging to a specified subset of N such as F.) Strong branching strategies[1] operate by selecting a set $F^o \subseteq F$, often focusing on a collection whose values $x_j^o$ are not close to integer values. Then, the approach solves the $2 \cdot |F^o|$ linear programs $(LP_j^+)$ and $(LP_j^-)$ derived from (LP⁰) by respectively setting $L_j^+ = \lceil x_j^o \rceil$ and $U_j^- = \lfloor x_j^o \rfloor$ for each $j \in F^o$; i.e., $L_j^+$ temporarily replaces the bound $L_j^o$ to produce (LP$_j^+$) and $U_j^-$ temporarily replaces the bound $U_j^o$ to produce (LP$_j^-$).

---

[1] Our comments related to strong branching are also generally applicable to reliability branching, which chiefly differs from strong branching by generating and relying on pseudo-costs at different junctures.

To compensate for the fact that the solution of the $2 \cdot |F^o|$ linear programs over $j \in F^o$ using strong branching can be computationally costly for any moderately large subset $F^o$ of $F$, preferred variants of such branching employ rules to allow early termination of the LP solution when the process has gone "far enough," chiefly by limiting the number of pivots allowed, and additionally using pseudo-costs in place of expensive LP solution steps after collecting information from past branches as a way to estimate the outcomes of current branches. We examine considerations related to both of these issues in the following sections.

### 2.1 Branching Strategies from Basic Considerations

Let Eval$_j^+$ denote the (objective function) evaluation for the up branch that sets $L_j^+ = \lceil x_j^o \rceil$, and let Eval$_j^-$ denotes the evaluation for the down branch that sets $U_j^- = \lfloor x_j^o \rfloor$. Commonly, reference is made to the fractional parts of $x_j^o$ given by $f_j^+ = \lceil x_j^o \rceil - x_j^o$ and $f_j^- = x_j^o - \lfloor x_j^o \rfloor$ and hence we may also write $L_j^+ = x_j^o + f_j^+$ and $U_j^- = x_j^o - f_j^-$. Viewing an up or down evaluation as a cost for moving away from the solution to (LP$^o$), and letting $x_o^o$, $x_{oj}^+$ and $x_{oj}^-$ denote the optimum $x_o$ values for (LP$^o$), (LP$_j^+$) and (LP$_j^-$), we may write

$$\text{Eval}_j^+ = x_{oj}^+ - x_o^o$$
$$\text{Eval}_j^- = x_{oj}^- - x_o^o$$

Later we propose more elaborate types of evaluations that may lead to better branching decisions and to improved rules for growing the tree by selecting among *open* tree nodes.

When using evaluation functions for branching, the choice of the winning variable $x_k$ is generally based on seeking a balance between the conflicting goals of maximizing and minimizing the smaller of Eval$_j^+$ and Eval$_j^-$. (Maximizing is important to assure that both branches will create an impact, and hence produce more equally balanced subtrees, by driving $x_o$ farther from its current value.) An often-used criterion proposed in Linderoth and Savelsberg (1999) is to take a convex combination of the two values Max$_j$ = Max(Eval$_j^+$, Eval$_j^-$) and Min$_j$ = Min(Eval$_j^+$, Eval$_j^-$), using a weight generally between 1/6 and 1/3 for Max$_j$.

However, a criterion that appears to work better is to take the product of Max$_j$ and Min$_j$ (Achterberg, Koch and Martin, 2005), which can be expressed as the following:

*Criterion 1.* Choose $x_k$, $k \in F^o$ to maximize Eval$_j$ = Eval$_j^+ \cdot$ Eval$_j^-$, $j \in F^o$.

A product term of 0 (or "near 0") in this criterion is replaced by a small positive quantity, recommended by Achterberg, Koch and Martin to be $10^{-6}$. Then the chosen $x_k$ is the source of an up branch if Eval$_k^+$ < Eval$_k^-$ and a down branch otherwise. (A branch that is infeasible automatically mandates the execution of the opposite branch, and the condition where both branches are infeasible compels the method to abandon the node, as by backtracking or seeking another open node in the tree.)

We propose additional criteria as alternatives to Criterion 1. For the first pair of these alternatives we adopt a perspective that favors generating fewer nodes before finding an optimal solution, as opposed to creating a B&B tree with fewer nodes overall. (Hence, when a problem is hard enough to that the solution process must be terminated before completing a full B&B tree, the motive is to have a higher probability of having found an optimal solution upon termination.) From this point of view, we suggest that Criterion 1 sometimes does not sufficiently differentiate between the values $Eval_j^+$ and $Eval_j^-$ to provide a useful means for determining which branch is preferable. The following two criteria are motivated with this in mind, the first of which comes in two forms, each based on a nonnegative parameter p which is an exponent for the term $|Eval_j^+ - Eval_j^-|$ (or equivalently, the term $Max_j - Min_j$):

*Criterion 2*. Choose $x_k$, $k \in F^o$ to maximize
  (a) $Eval_j = Eval_j^+ \cdot Eval_j^- \cdot |Eval_j^+ - Eval_j^-|^p$, $j \in F^o$.
  (b) $Eval_j = Min_j \cdot |Eval_j^+ - Eval_j^-|^p$, $j \in F^o$.

Once again, a 0 product term is replaced by a small positive value.

Note when p = 0 Criterion 2(a) becomes the same as Criterion 1, while for p sufficiently large Criteria 2(a) and 2(b) both reduce to maximizing $|Eval_j^+ - Eval_j^-|$. Two variants of the foregoing that likewise invite exploration result by replacing $|Eval_j^+ - Eval_j^-|^p$ in (a) and (b) by $Max_j^p$, though of course the best value for p will then change. (For example, the best p may be less than 1 for $|Eval_j^+ - Eval_j^-|^p$ and greater than 1 for $Max_j^p$,)

For our third criterion, define $MinMin(F^o) = Min(Min_j: j \in F^o)$ and $MaxMin(F^o) = Max(Min_j: j \in F^o)$. The criterion is based on selecting a value for a parameter $\lambda \in [0,1]$ to yield a threshold $T(\lambda) = MinMin(F^o) + \lambda(MaxMin(F^o) - MinMin(F^o))$ and then to compel $Min_j \geq T(\lambda)$. Evidently, a value of $\lambda$ closer to 1 yields a threshold that admits only the larger $Min_j$ values (which fall closer to $MaxMin(F^o)$).

*Criterion 3*. Choose $x_k$, $k \in F^o$ to maximize $Eval_j = |Eval_j^+ - Eval_j^-|$, $j \in F^o$ subject to $Min_j \geq T(\lambda)$.

As a crude approximation, Criterion 3 with $\lambda = .9$ will give results roughly similar to those of Criterion 1, and Criterion 3 with $\lambda = .75$ will give results more nearly similar to those of Criterion 2(a) or 2(b) in the case where p = 1. We note that applying Criterion 3 with larger values of $\lambda$ can be useful for the goal of keeping the unexplored portion of the tree – the portion that is rooted by the lower evaluation node, which is not currently selected – as small as possible.

## 2.2 Branching Strategies and Evaluation Criteria from MIP Applications in Simulation Optimization

Our next criteria for choosing a winning variable $x_k$ to branch on come from experiments with solving MIP problems encountered in the simulation optimization setting, using the MIP software coded for the OptQuest system (www.OptTek.com). Although outcomes from problems in this setting may not be indicative of those to be expected more generally, the findings from these experiments may be useful for suggesting alternatives that may be relevant for special cases. Each of the following two criteria were found to work well for MIP applications arising in the simulation optimization context, by taking $F^o = F$, the full set of fractional variables.

*Criterion 4.* Choose $x_k$, $k \in F^o$ to maximize $Eval_j = Max_j \cdot |Eval_j^+ - Eval_j^-|$, $j \in F^o$.

The next criterion, like Criterion 2 earlier, depends on a nonnegative parameter p.

*Criterion 5.* Choose $x_k$, $k \in F^o$ to maximize $Eval_j = Min_j^p \cdot (Eval_j^+ + Eval_j^-)$, $j \in F^o$.

A good value for p suggested by previous experiment is $p = .3$, though more extensive experimentation in other settings is likely to uncover a better value.

The most effective criterion from the simulation optimization setting takes a somewhat different form that changes the evaluation criteria to include reference to a weighted sum of integer infeasibilities and a second weighted sum of constraint infeasibilities. (The latter is incorporated in cases where LP problems are not solved all the way to feasibility by the dual method. Such cases are relevant in applying the Winnowing Procedure discussed later.) Let $F^+(j)$ and $F^-(j)$ denote the sets of fractional variables for the problems $(LP_j^+)$ and $(LP_j^-)$, and denote the fractional values that correspond to $f_i^+$ for the problem $(LP^o)$ by $f_i^+(j)$, $i \in F^+(j)$ and $f_i^+(j)$, $i \in F^-(j)$. (It is to be understood that the values $f_i^+(j)$ (and their associated values $f_i^-(j)$) differ according to whether $i \in F^+(j)$ or $i \in F^-(j)$. Our notation is chosen to avoid more complex designations such as $f_i^{++}(j)$ and $f_i^{+-}(j)$, and $f_i^{-+}(j)$ and $f_i^{--}(j)$.) Also, let $Infeas^+$ and $Infeas^-$ respectively denote the sum of constraint infeasibilities expressed as positive quantities (including bound violations) for the problems $(LP_j^+)$ and $(LP_j^-)$, which are taken from a current dual basic solution that is not primal feasible. Then for given positive weights $w_1$ and $w_2$ we define

$$(D1)\ Eval_j^+ = x_{oj}^+ - x_o^o + w_1(\sum Min(f_i^+(j), f_i^-(j)): i \in F^+(j)) + w_2 \cdot Infeas^+$$
$$(D2)\ Eval_j^- = x_{oj}^- - x_o^o + w_1(\sum Min(f_i^+(j), f_i^-(j)): i \in F^-(j)) + w_2 \cdot Infeas^-$$

The values $w_1$ and $w_2$ depend on rules employed for scaling the objective function and constraints, but $w_1 = 100$ and $w_2 = 10$ worked well for the scaling rules applied in the simulation optimization setting. The disproportionately large emphasis on achieving integer feasibility indicated by the value of $w_1 = 100$ may be due to the fact that the MIP method in this instance operated by imposing optimistic bounds on $x_o$ (that were successively relaxed, when necessary, by a process that recovered relevant portions of the B&B tree). Assigning a value to $w_2$ of course becomes superfluous when strong branching is employed and LP problems are solved all the way to optimality (hence achieving feasible solutions). Experimentation can undoubtedly determine better $w_1$ and $w_2$ values for problems in other settings.

When (D1) and (D2) are used for branching decisions, they can evidently be embedded in any of the choice criteria previously described. In addition, the following somewhat different approach was taken in the context where these definitions were introduced. (Again, this criterion takes $F^o = F$.)

*Criterion 6*. Choose $x_k$, $k \in F^o$ to minimize $Min_j$, $j \in F^o$, for $Min_j = Min(Eval_j^+, Eval_j^-)$ defined by reference to (D1) and (D2).

This choice of the minimum of the $Min_j$ quantities reflects the fact that, upon ultimately choosing a variable $x_k$ to branch on, the minimum of the values $Eval_k^+$ and $Eval_k^-$ is the one that determines whether $x_k$ will preferably branch up or down. The incorporation of weighted infeasibilities in Criterion 6 makes the focus on this minimum more reasonable than for other definitions of $Eval_k^+$ and $Eval_k^-$ indicated earlier, though the option of incorporating (D1) and (D2) in other criteria indicated earlier should not be overlooked.

As our final criterion, we propose an extended variant of Criterion 6 which replaces the $Min(f_i^+(j), f_i^-(j))$ terms in (D1) and (D2) with terms that more closely capture the objective function impact of branching with respect to these fractional values. Here, we make additional use of the branching costs $x_{oj}^+ - x_o^o$ and $x_{oj}^- - x_o^o$ for the up and down branches for $x_j$, by identifying the associated unit costs

$$UC_j^+ = (x_{oj}^+ - x_o^o)/f_j^+ \text{ and } UC_j^- = (x_{oj}^- - x_o^o)/f_j^-, j \in F^o.$$

In our development, it is understood that a positive epsilon value is inserted in place of the numerator if the numerator is 0.

These unit costs are the same ones customarily incorporated into pseudo-costs, as discussed in Section 5, but we employ them in a different way.[2] Having first determined $UC_j^+$ and $UC_j^-$ by solving the problems $(LP_j^+)$ and $(LP_j^-)$ for each $j \in F^o$, we then apply them to obtain costs implicitly associated with the fractional values $f_i^+(j)$ and $f_i^-(j))$ for $i \in F^+(j)$ and $i \in F^-(j)$. Since $F^+(j)$ and $F^-(j)$ may each contain fractional variables not contained in $F^o$, it may be that $UC_i^+$ and $UC_i^-$ are not defined for some indexes $i \in F^+(j)$ or $i \in F^-(j)$. To handle this, we define $UC_j = |RC_j|$ for $j \in N - F^o$, where $RC_j$ is the reduced cost for variable $x_j$ in the dual simplex solution to $(LP^o)$ (where $RC_j = 0$ for basic variables). Then we define

$$MinCost_i^+ = Min(UC_i^+ \cdot f_i^+(j), UC_i^- \cdot f_i^-(j)): i \in F^+(j)$$
$$MinCost_i^- = Min(UC_i^+ \cdot f_i^+(j), UC_i^- \cdot f_i^-(j)): i \in F^-(j)$$

---

[2]. In particular, we focus here on evaluations taken directly from tentative branches at a node which is a parent of the nodes where these evaluations are applied. The relevance of this modified focus will be clarified in Section 5, where we introduce modifications of the customary pseudo-cost approaches based on a similar idea.

We emphasize again that the quantities $f_i^+(j)$ and $f_i^-(j)$ for $i \in F^+(j)$ differ from those for $i \in F^-(j)$. However, $UC_i^+$ and $UC_i^-$ depend only on the index $i$, independent of the set $F^+(j)$ or $F^-(j)$ that contains this index. Also $MinCost_i^+$ and $MinCost_i^-$ are positive as long as the associated branch is not MIP feasible, by the "epsilon convention" that assures the values $UC_i^+$ and $UC_i^-$ are positive.

The idea underlying our use of $MinCost_i^+$ and $MinCost_i^-$ is that a more informed type of evaluation results by considering the "second order" objective function effect of implicitly branching on variables $x_i$ that are fractional in the solutions produced by the $x_j$ up and down branches. We seek to capture this second order effect by means of the evaluations

$$(D3) \; Eval_j^+ = x_{oj}^+ - x_o^o + w_1(\sum MinCost_i^+ : i \in F^+(j)) + w_2 \cdot Infeas^+$$
$$(D4) \; Eval_j^- = x_{oj}^- - x_o^o + w_1(\sum MinCost_i^- : i \in F^-(j)) + w_2 \cdot Infeas^-$$

The weights $w_1$ and $w_2$ will have different preferred values here than in (D1) and (D2). We observe that the summations over $MinCost_i^+$ and $MinCost_i^-$ imperfectly capture the manner in which these approximating terms impact the objective function, since their effects are not additive. A potentially useful way to compensate for the non-additivity effect would be to include only a limited number of the largest $MinCost_i^+$ and $MinCost_i^-$ terms in their respective summations, or more generally to assign decreasing weights (that eventually become 0) as these terms grow smaller.[3] However, we hypothesize that (D3) and (D4) can prove useful without such refinements.

A choice criterion that results by extension of Criterion 6 may then be expressed as.

*Criterion 7.* Choose $x_k$, $k \in F^o$ to minimize $Min_j$, $j \in F^o$, for $Min_j = Min(Eval_j^+, Eval_j^-)$ defined by reference to (D3) and (D4).

The unit cost terms $UC_i^+$ and $UC_i^-$ incorporated into the definitions of $MinCost_i^+$ and $MinCost_i^-$ normally require solving the problems $(LP_j^+)$ and $(LP_j^-)$ to optimality (to obtain feasible solutions) in order for these costs to be fully meaningful. Hence, in the usual case where the LP solutions to $(LP_j^+)$ and $(LP_j^-)$ are feasible, the use of $w_2$ in (D3) and (D4) is superfluous. An exception may occur in applying the Winnowing Procedure of the next section, which employs a screening stage that may not completely solve the LP problems for the branches considered.

---

[3] A similar approach to account for non-additivity could be employed in the summations of (D1) and (D2). A more ambitious approach could be undertaken by tracking interactions among variables, to note which branches of particular variables cause other variables to change their values by various degrees of magnitude in specific directions.

If the definitions of $Eval_j^+$ and $Eval_j^-$ from (D3) and (D4) are embedded in one of the earlier criteria instead of being used with Criterion 7, a natural alternative would be to employ (D3) and (D4) within Criterion 3, in conjunction with a small value of $\lambda$. We also observe the possibility of using more than one of the criteria proposed above by combining them to vote on the best branching alternative (including conditional voting, as where a first step isolates a subset of alternatives which are then subjected to a final vote).

### 2.3 Using the Evaluation Criteria for Selecting an Open Node in the B&B Tree.

The evaluations $Eval_j^+$ and $Eval_j^-$ given by (D1) and (D2), or alternatively by (D3) and (D4), can be used in a natural way for the purpose of selecting an open node in the B&B tree. Let $Eval_j^\alpha$ denote $Eval_j^+$ or $Eval_j^-$ (interpreting $\alpha$ as + or -) according to the whether the open node was generated by an up or down branch. We assume that the "heuristic accuracy" of $Eval_j^\alpha$ generally improves as the depth of the node in the tree increases. (Clearly, if the node represents a feasible MIP solution the accuracy of the evaluation is perfect.)

More precisely, we seek an adjustment (calibration) of these evaluations to produce a depth-based function $Dval_j^\alpha(d)$ of $Eval_j^\alpha$ applicable to nodes that lie at depth d from the root (where the depth of the root is $d = 0$, and by implication all open nodes have $d \geq 1$). One approach is to determine $Dval_j^\alpha(d)$ so that it will (approximately) match the value $x_o^* - x_o^o$, where $x_o^*$ is the $x_o$ value of the best MIP feasible descendant of the current node and $x_o^o$ is the $x_o$ value at the node to which $Eval_j^\alpha$ applies.

We refer to the definitions of (D3) and (D4), which also encompass the customary evaluation definitions by choosing $w_1 = w_2 = 0$ and encompass the definitions of (D1) and (D2) by defining the MinCost values appropriately (by reference to $Min(f_i^+(j), f_i^-(j))$). For simplicity we assume the Infeas values are 0 for open nodes (i.e., the LP problem has been solved to optimality, or very nearly so) in contrast to tentatively generated nodes that may be produced by winnowing procedures as described subsequently. Consequently the weight $w_2$ can be considered to be 0 and (D3) and (D4) may be summarized by the following expression (for $\alpha = +$ or -):

$$(D0)\ Eval_j^\alpha = x_{oj}^\alpha - x_o^o + w_1(\sum MinCost_i^\alpha: i \in F^\alpha(j))$$

A simple way to define the function $Dval_j^\alpha(d)$ of $Eval_j^\alpha$ (by reference to (D0)) is to select a weight $w(d)$ based on d to give

$$Dval_j^\alpha(d) = w(d) \cdot Eval_j^\alpha$$

Then to match $Dval_j^\alpha(d)$ to $x_o^* - x_o^o$ we simply define

$$w(d) = (x_o^* - x_o^o)/Eval_j^\alpha$$

Such a determination of course depends on appropriately selecting nodes whose evaluations $Eval_j^\alpha$ can be used to assign a value to $w(d)$. To obtain a first set of candidate values for $w(d)$ (apart from temporarily relying on default values from previous experience), let $x^*$ denote the first feasible MIP solution found (which is the best solution known when it is identified), yielding $x_o = x_o^*$, and let $d^*$ denote the depth of $x^*$ as a node of the B&B tree. Then we select the nodes that lie on the unique path from the root to $x^*$ to be the source of the $x_o^o$ and $Eval_j^\alpha$ values used to define $w(d) = (x_o^* - x_o^o)/Eval_j^\alpha$ for $d = 1$ to $d^* - 1$.

Note that for the customary definition of $Eval_j^\alpha$ from the literature (in which $w_1 = 0$) we have $w(d) = (x_o^* - x_o^o)/(x_{oj}^\alpha - x_o^o)$, identifying $x_o^o = x_o^o(d)$ for the solution $x^o$ at node $d$. Assuming the LP problems for the nodes along the path are always solved to optimality, we therefore have $x_{oj}^\alpha = x_o^o(d+1)$. The numerator $(x_o^* - x_o^o)$ thus decreases or stays the same as $d$ gets larger.

A more general expression for $Dval_j^\alpha(d)$ occurs by introducing two weights, $w_o(d)$ and $w_1(d)$, where $w_o(d)$ replaces $w(d)$ as a weight for $Eval_j^\alpha$ above, and $w_1(d)$ replaces $w_1$. Assume the values $x_{oj}^\alpha - x_o^o$ and $MinCost_i^\alpha$ have been saved along this path from the root to $x^*$. For the node on this path at $d = d^* - 1$, the appropriate value for $w_o(d)$ is 1 and the value of $w_1(d)$ is irrelevant since $MinCost_i^\alpha = 0$ at this node. Thus, we are interested in compute $w_o(d)$ and $w_1(d)$ for $d$ satisfying $d^* - 1 > d \geq 1$.

**Approach 1.**
Simplest is to fix $w_o(d)$ at the constant value $w_o = 1$ for all $d$ and allow $w_1(d)$ absorb all the necessary variation to define $Dval_j^\alpha(d)$ appropriately. Then, using $w_o(x_{oj}^\alpha - x_o^o) + w_1(\sum MinCost_i^\alpha: i \in F^\alpha(j)) = (x_o^* - x_o^o)$, the value of $w_1 = w_1(d)$ is simply given by

$$w_1 = (x_o^* - x_{oj}^\alpha)/(\sum MinCost_i^\alpha: i \in F^\alpha(j)).$$

The denominator must be positive by conventions noted earlier since the nodes for $d^* - 1 > d \geq 1$ are not MIP feasible. In the case of degeneracy, where $x_{oj}^\alpha$ may equal $x_o^*$ in spite of resulting from a solution that is not MIP feasible, then $w_1 = 0$ and the resulting estimator $Dval_i^\alpha$ for matching $x_o^* - x_o^o$ is given simply by setting $w_o = 1$ to yield the simplest form of $Eval_i^\alpha = x_{oj}^\alpha - x_o^o$.

Having thus determined $w_o(d) = 1$ and $w_1(d)$ as specified for each value of $d$ satisfying $d^* - 1 > d \geq 1$, we extend these values for $d \geq d^* - 1$ by continuing to set $w_o(d) = 1$ and setting $w_1(d) = w_1(d^* - 2)$. If $d^* - 2 < 1$, i.e. $d^* = 1$ or $2$, we observe in the former case that the MIP feasible solution was found as a child node of the root of the B&B tree, and in the latter case that it was obtained as grandchild, and in either case we use the degenerate default of $w_o = 1$ and $w_1 = 1$.

However, once additional MIP feasible solutions are found we modify the $w_1(d)$ values by computing the values for the new solution and then taking the average over all values achieved to date.

**Approach 2.**
A second method for determining $w_o(d)$ and $w_1(d)$ is to derive them from the solution of the two simultaneous equations

$$w_o(x_{oj}^\alpha - x_o^o(d)) + w_1(\sum MinCost_i^\alpha : i \in F^\alpha(j)) = x_o^* - x_o^o(d)$$
$$w_o(x_{oj}^\beta - x_o^o(d+1)) + w_1(\sum MinCost_i^\beta : i \in F^\beta(j)) = x_o^* - x_o^o(d+1)$$

where $x_o^o(d)$, $x_{oj}^\alpha$ and $MinCost_i^\alpha$ refer to the indicated values associated with the node at depth d, and where $x_o^o(d+1)$, $x_{oj}^\beta$ and $MinCost_i^\beta$ refer to these values at depth $d + 1$. (The index j of course differs according to the case.)

The solution to these equations gives the values for $w_o(d)$ and $w_1(d)$. If $d = d^* - 2$ (the largest d previously considered) then $d + 1 = d^* - 1$, which implies $x_{oj}^\beta = x_o^*$ and $F^\beta(j)$ is empty. Hence $w_o = 1$ from the second equation, and then the value determined for $w_1$ ($= w_1(d)$) is the same as in Approach 1 where $w_o$ is uniformly assigned the value 1. Thus, once again, we determine $w_o(d)$ and $w_1(d)$ for all values of d satisfying $d^* - 1 > d \geq 1$, except that now $w_o(d)$ can vary across different d values. As before, when new MIP feasible solutions are found we simply average their $w_o(d)$ and $w_1(d)$ values with those previously obtained to get the "working values" used to evaluate open nodes to determine which one is selected next.

With this background, we are prepared to go beyond the consideration of evaluation criteria to address the principal focus of this paper.

## 3. Narrow Gauge Branching Strategies

The strong branching and reliability branching approaches to single out a particular variable that appears most attractive for branching may be viewed as *broad gauge* strategies in the sense that they solve, or approximate the solution of, linear programs for a set F of variables that can be relatively large (particularly in problems of large dimension). Relative to the LP problems solved, these strategies are therefore "wide and shallow."

We continue to recognize the importance of branching on a variable $x_k$ that has a relatively high impact on the current solution to (LP$^o$),[4] but are also motivated to emphasize the relevance of branching in the right direction, which is to say, the direction that leads to an optimal MIP solution. (This motivation underlies several of the new criteria of Section 2, as previously intimated.) We are additionally influenced by the fact that it is only necessary to identify this "right decision" for a single variable in F in order to progress toward optimality. Consequently, we propose a *narrow gauge* strategy that confines the candidate set $F^o \subset F$ of fractional variables for branching to be substantially smaller than normally considered. More specifically, by a narrow gauge approach we mean a method that examines variables from this restricted candidate set to greater depths than by simply solving the single LP for each branching alternative – hence accounting for fuller ramifications of the current branching alternatives.

---
[4] The policy of branching to force change is shown to be particularly valuable in Pryor and Chinneck (2011), whose use a rule to make the children far from the parent using a Euclidean distance measure.

In sum, instead of employing a single "LP look-ahead branching" for a fairly large number of variables as in strong branching and reliability branching, we propose an approach that creates an "LP look-ahead tree" for a small number of variables. The look-ahead tree may be generated completely to a specified depth or, by rules subsequently indicated, may be pruned by removing or disregarding certain branches.

An instance of such a strategy was proposed by Glankwamdee and Linderoth (2011) for a look-ahead approach that considers options one step beyond the customary depth 1 level for evaluating choices. Our approach differs from theirs in the following respects. First, we employ special winnowing procedures for isolating branching variables that make it cost effective to generate look-ahead trees of greater depth. Second, drawing on the ideas of the previous section, we determine the branch choices at intermediate levels, and ultimately infer the preferred branching variable and branching direction at the initial level, by criteria than differ from those employed for making such choices in the previous literature. Third, we introduce a collection of strategies that further alter the tree structure and the decisions ultimately made. Fourth, we introduce mechanisms for branching on derivative variables to refine the information generated and the bounds produced. Finally, we provide complementary procedures in later sections of the paper that enlarge the scope of the narrow gauge approach.

To obtain a rough idea of the tradeoff between narrow gauge and customary broad gauge strategies, consider a small look-ahead tree of depth 3. Such a tree generates $2^3$ leaf nodes or a total of $2^1 + 2^2 + 2^3 = 14$ nodes in all (excluding the node for the current root at (LP$^o$).) An additional computational cost is incurred in order to select the branches leading to these nodes, as will be shown, which effectively increases the computation by an amount equivalent to generating in the vicinity of 64 to 74 nodes in total. Consequently, the solution time for generating the tree would be roughly comparable to that of a strong branching process that examines a set of 32 to 37 candidate variables (which produces 2·32 to 2·37 branches to examine). We hypothesize that the additional information made available by this narrow gauge approach for the purpose of choosing an initial (up or down) branch is sufficiently valuable to prefer looking deeper in place of examining a larger number of initial branching variables.

**An Idealized Example.**
To clarify the rationale behind the use of a look-ahead tree in the narrow gauge approach, consider an idealized situation where computational resources are available to examine the same number of branching alternatives at each node of the narrow gauge tree as examined by strong branching. To generate the 14 nodes of a tree of depth D = 3, we would execute strong branching at each node for depths d = 0 to d = 2, since this will result in identifying the best (highest evaluation) branching variable for each node at d = 3, and thus produce the 8 leaf nodes at d = 3.

The number of nodes where strong branching occurs at the depths d = 0 to 2 is $1 + 2 + 2^2 = 7$, hence causing this tree-based modification of strong branching to consume 7 times the effort of ordinary strong branching. Alternatively, instead of using this computation to choose only the branching variable at depth d = 0, we could employ a variant that accepts the entire path to the winning node at depth d = 3, thus reducing overall computation to a factor of 7/3 = 2 1/3 times the amount used by ordinary strong branching.

Within a scenario that allows such an increase in computational effort, suppose that the use of strong branching in general produces a .6 probability that a correct branch will receive an evaluation that favors its selection in preference to that of any incorrect branch. Then there is a .84 probability of encountering such an evaluation at least upon reaching depth 2 of the tree, and a .936 probability of encountering it upon reaching depth 3. This example of course is greatly oversimplified and ignores a number of relevant considerations, but provides a sense of the potential utility of the tree-based approach employed in the narrow gauge strategy.

We will return again to this idealized example under the topic of Persistent Attractiveness in Section 7.1.

## 4. Generating Branches of the Narrow Gauge Tree.

We begin by introducing a *progressive winnowing* strategy to reduce the computational effort of generating the branches of the narrow gauge look-ahead tree, and then describe the branch generation process. A number of variations of the approach are indicated subsequently.

### 4.1 Progressive Winnowing

Progressive winnowing applies a succession of screening rules to isolate fewer and more promising candidates at each stage. Starting from an initial stage that applies a relatively weak rule for segregating good choices from bad ones, the approach applies stronger and stronger rules in successive stages until the number of candidates is reduced to a preferred number. We use the term *Specified Criterion* to refer to a selected instance of one of the criteria of Section 2. The parameters of the process ($n_0$, $n_1$, $n_2$ and an associated quantity $k_2$ below) are to be determined by experimentation.

**Winnowing Procedure**
Stage 1. Employ preliminary screening to create a set $F_0 \subset F$ by selecting $n_0$ variables $x_j$, $j \in F$, with values $f_j^-$ closest to .5 (equivalently with $|x_j^o - .5|$ closest to 0). (The value $n_0$ will usually be chosen large enough to admit all variables in F as candidates, hence yielding $F_0 = F$.) Then select $n_1$ of the $n_0$ variables in $F_0$ to create a set $F_1 \subset F_0$, where the Specified Criterion is applied to evaluations $Eval_j^+$ and $Eval_j^-$ derived from executing a single dual pivot.[5] The value $n_1$ may be calibrated to be a fraction of the average number of fractional variables in F.

Stage 2. Create a set $F_2 \subset F_1$ by selecting $n_2$ of the $n_1$ variables from Stage 1 that yield the largest evaluations by the Specified Criterion, where $Eval_j^+$ and $Eval_j^-$ are determined by branching on $x_j$ for $j \in F_1$ and performing $k_2$ dual simplex pivots on each branch. The value $k_2$ may be calibrated as a fraction of the average number of pivots normally used to solve a problem (LP°) to optimality.

*Calibrating $n_1$ and $k_2$*: A concrete example will help to give a sense for calibrating $n_1$ and $k_2$. Suppose that F contains an average of 40 fractional variables (which falls below the limit of 100 such variables normally permitted for consideration) and that the average number of pivots used to solve a problem (LP°) to optimality is 150 (which falls inside the 1000 upper limit on pivots typically allowed in strong branching)[6]. Further suppose $n_1$ is selected to be 1/4 the 40 average size of F, and $k_2$ is selected to be 1/6 the 150 average number of pivots for solving (LP°). This means we examine 10 (= 40·1/4) additional pairs of branches (20 nodes) for each node of the narrow gauge tree where winnowing is applied, but perform 1/6 of the computation for each of these added nodes that would be carried out by strong branching, and hence add 20/6 = 3 1/3 "nodes worth" of computation to each node. For a tree of depth D = 3, the Winnowing Procedure is applied for d = 0, 1 and 2, hence increasing computation by an amount equivalent to adding 3 1/3 nodes to each of the 7 (= 1 + 2 + 4) nodes generated. Thus the total effort amounts to adding roughly 24 nodes to the 30 of the depth D = 3 tree (rounding up to account for the slight additional effort of the single dual pivot strategy in Stage 1).

Next we consider the effect of the value given to $n_2$ in Stage 2.

*Selecting $n_2$ in Stage 2*: The value of $n_2$ determines the number of branching variables whose corresponding LP problems will be solved to optimality in at each node of the narrow gauge tree in order to choose the up and down branching variable at the node. In reality, $n_2$ depends on the value of d, because we place greater emphasis on solving (a few) LP problems to optimality when d = 0 than at greater depths. To illustrate, the value of $n_2$ may be selected to lie in the range from 3 to 8 for d = 0, and be restricted to be only 2 for d = 1. (thus exploring 4 branching variables in total for the two nodes generated at d = 1). Thereafter, it is reasonable to select to $n_2 = 1$ for all greater depths d > 1, hence accepting the branching variable evaluated as best by Stage 2 of the Winnowing Procedure and solving a single LP problem for each branch. Thus, in sum, the 6 to 16 (= 2·3 to 2·8) added nodes from d = 0, together with the 4 (= 2·2) added nodes for d = 1 join with the approximately 54 nodes previously mentioned to give a computation equivalent to examining between 64 and 74 branches (nodes) – 32 to 37 branching variables – with strong branching.

---
[5] This is motivated by the fact that up and down branching costs for a single dual pivot can be computed with almost negligible effort; i.e., the change in $x_o$ can be computed from ratios of reduced costs to pivot row coefficients, without bothering to execute the pivot itself. For an up branch, these ratios are negative or positive, respectively, for nonbasic variables at their lower or upper bounds. For a down branch, the opposite is true. Then the minimum absolute value of these ratios is multiplied by $f_j^+$ or $f_j^-$, respectively, to give the amount by which $x_o$ increases as a result of the pivot. Two-pivot evaluations, which require somewhat more effort, can capitalize on similar streamlining once the second pivot row is selected, and hence may also be relevant to consider.

[6] Achterberg, Koch and Martin (2005).

The illustrated values of the parameters $n_1$, $k_2$ and $n_2$ can be adjusted to change the total computation as desired. For example, the number of pivots $k_2$ may be dropped to somewhere between 5 and 15 if appropriate weights $w_1$ and $w_2$ in Criterion 6 and 7 of Section 2 are used to help capture the consequences of not completing the LP solution process. Of course we can alternatively employ the variation mentioned in the Idealized Example of Section 3 to accept a path to a leaf node of the tree rather than to use the tree to select only the first branch, thus effectively reducing the total computation by a factor of D, dividing the effort by 3 in the case of the example given just above. In addition, we later examine variants which reduce the number of nodes and branches in the narrow gauge tree, which can result either in increasing D or permitting larger values of $n_2$ to be selected (especially at $d = 0$ and $d = 1$).

Drawing on the Winnowing Procedure, we now explicitly describe the narrow gauge strategy applied to the generation of a single tree, and afterward indicate a simple modification for generating more than one tree.

### 4.2 Narrow Gauge Strategy to Generate a Single Look-ahead Tree

The current problem ($LP^o$) represents the tree node at depth $d = 0$ as and we scan tree nodes from depth $d = 0$ to depth $d = D – 1$, where D is the maximum depth of the tree. Each node scanned at depth d generates 2 child nodes at depth $d + 1$. We indicate the character of the tree as generated by a breadth first process, which allows the simplest description.

**Basic Tree Generation Framework (Breadth First Format)**
*Begin tree generation*
For $d = 0$ to $D – 1$
    For each tree node at depth d, generate two child nodes at depth $d + 1$:
        (1) Apply the Winnowing Procedure to the current node to generate the set $F_2$.
        (2) Generate and solve the two LP problems ($LP_j^+$) and ($LP_j^-$), for each $j \in F_2$ and
           apply the Specified Evaluation Criterion to select a single branching variable
           $x_h$, $h \in F_2$ thus adding two nodes to the look-ahead tree at depth $d + 1$.
    End generation of nodes at depth $d + 1$
End iterations over depth $d = 0$ to $D – 1$
(3) Evaluate each pair of sibling nodes at depth D (each pair of leaf nodes that share a common parent), by defining $Eval_j^+ = (x_o^+ – x_o^o)$ and $Eval_j^- = (x_o^- – x_o^o)$ using Criterion 2 or Criterion 3. (Here $x_o^+$ and $x_o^-$ refer to the $x_o$ values for the problems ($LP_h^+$) and ($LP_h^-$) at depth D, and $x_o^o$ refers to the $x_o$ value for the problem ($LP^o$) associated with $d = 0$.)
(4) Identify the initial branch at node 0 that generates the subtree containing the selected pair. This branch gives the current up or down branch at node 0 to be selected for $x_k$, and the resulting node at depth 1 on this branch becomes the new node 0.
*End tree generation*

While the foregoing breadth first description is useful for simplicity, we emphasize that the resulting structure should preferably be produced by a depth first process, which provides a more economical use of memory.

*Comment 1*: The process of solving the LP problems $(LP_j^+)$ and $(LP_j^-)$ in Step (2) may be terminated after a standard limit on the number of allowable pivots (e.g., 1000) or after a limit on the number of pivots without changing the value of $x_o$ in the case of dual degeneracy. There are two special cases to keep in mind: (i) If any of the LP problems $(LP_j^+)$ or $(LP_j^-)$ is determined to have no feasible solution at any point of the Winnowing Procedure in Step (1), or at any point during Step (2), then as previously noted the complementary problem represents a compulsory branch which is immediately executed to modify $(LP^o)$ (and the tree generation process begins again). Likewise, if both branches $(LP_j^+)$ and $(LP_j^-)$ are infeasible, the node for $(LP^o)$ is itself classified as infeasible and is abandoned. We note that a problem $(LP_j^+)$ or $(LP_j^-)$ can be considered infeasible if at any dual iteration $x_o \geq x_o^*$, where $x_o^*$ denotes the objective function value for the best solution $(x^*, y^*)$ found. (ii) If the solution to the LP problem at a given node is MIP feasible, then it is automatically a new best solution found since applying the criterion of (i) that treats an LP problem as infeasible when $x_o \geq x_o^*$ assures $x_o < x_o^*$. Upon recording the new best solution and updating $x_o^*$, the current LP node that produced this new best solution becomes infeasible since now $x_o = x_o^*$, and hence the alternative branch becomes compulsory.

*Comment 2*: When a new best solution is found, the new $x_o^*$ value may imply the infeasibility of nodes previously generated in the tree, and such nodes are then removed.

*Comment 3*: We have specified that Criterion 2 or Criterion 3 be applied in Step (3) in the expectation that these criteria will be preferable to Criterion 1. That is, we seek a stronger rule than Criterion 1 for differentiating the quality of the sibling nodes at depth D that will determine the up or down branch selected for $x_k$ in Step (4). (Note that when the tree is generated by depth first search, sibling leaf nodes will be generated consecutively, enabling their evaluations to be immediately compared.) A modified version of these criteria may also be applied that places a weight on the number of fractional variables in the solutions at the leaf nodes, or on the "fractionality" of these solutions, i.e., the sum of the values $Min(f_j^+, f_j^-)$ over the fractional variables. (The weight should be negative in the context of maximizing the resulting evaluation.)[7]

*Comment 4*: If one member of a sibling node pair is missing at depth D (because the LP solution at this node is infeasible either by (i) or (ii) of Comment 1 above) then we treat the $x_o$ value for the missing child node as having $x_o = x_o^*$ even if the infeasibility did not result by finding a new best solution. This avoids considering the surviving child as unduly attractive compared to its missing sibling.

---

[7] As indicated previously, such a rule produced the best results for the OptQuest MIP system. We discuss variations in Section 6 when treating the subject of pseudo-cost evaluations.

*Comment 5*: We can generate more than one look-ahead tree by modifying Step (2) for the case d = 0 by and selecting a small number n' < $n_2$ of variables $x_h$, h ∈ $F_2$, with highest evaluations to be initial branching variables, each for its own tree. To prevent different trees from generating the same branches (in a different order), it suffices to order the initial branching variables $x_h$ for d = 0 as $x_1, x_2, \ldots, x_{n'}$, while giving all other variables higher indexes, and then require that the tree rooted at $x_h$ (= $x_1, x_2, \ldots,$ or $x_{n'}$) cannot include a branch (up or down) on any variable $x_j$ such that j < h. (This rule effectively imposes a classical lexicographic ordering for avoiding duplications.)

We now describe ways to enhance the foregoing basic tree generation approach.

### 4.3 Post-Winnowing

A natural modification of the narrow gauge strategy is a *post-winnowing* approach that intervenes in the construction of the trees so that, beginning at some selected depth $d_o$, the number of branching variables (hence paired sibling nodes) carried forward will be restricted to a limiting value Lim(d).

This approach may be implemented within the Basic Tree Generation Framework by inserting the following instructions immediately after the conclusion of the loop indicated by the statement "End generation of nodes at depth d + 1":

(2a) Retain only the Lim(d) pairs of sibling nodes at depth d + 1 whose branching variables $x_h$ yield the highest evaluations by the Specified Evaluation Criterion.
(2b) (Best sibling option) From each pair of sibling nodes retained in (2a), retain only the node that produced the smaller evaluation $Eval_j^+$ or $Eval_j^-$.
(2c) (Best single stream option) Across all pairs of sibling nodes generated at depth d + 1, select the Lim(d) nodes that have the best evaluations.

Reasonably, $d_o$ may be selected to be at least 2, to limit the number of sibling node pairs retained in going from depth d = 2 to depth d = 3, since this permits four pairs of sibling nodes to be the source of those chosen to be retained, while only 2 pairs are generated in going from d = 1 to d = 2.

To illustrate the application of (2a) without including the "Best sibling option" of (2b), consider the case where $d_o = 2$ and Lim(d) is given a constant value of Lim = 3. (Since Lim = 3 is greater than the number of sibling node pairs generated until reaching $d_o = 2$, it would be unnecessary to stipulate the value of $d_o$ in this case.)

Then the 3 highest evaluation branching variables (and their associated sibling nodes) from the 4 generated for going from d = 2 to d = 3 are chosen for continuing the tree at depth d = 3. Thus the tree will have just 6 leaf nodes at d = 3, and each of these will now generate 6 new pairs of sibling nodes in going from d = 3 to d = 4 (hence 12 leaf nodes at d = 4). From these 6 pairs we again retain the Lim = 3 best. Then once more we have 6 leaf nodes at d = 4 to carry forward, and we have entered a "steady state" situation in which at every depth $d > d_o$ we go from 3 pairs of leaf nodes to 6 pairs, which are then once more winnowed down to only 3 pairs. The number of nodes (LP problems) generated and examined in this process for D = 6 is 2 + 4 + 6 + 12 + 12 + 12 = 48 versus 2 + 4 + 8 + 16 + 32 + 64 = 126 for D = 6 (and 62 for D = 5) in the ordinary case. In this example we are ignoring the generation of additional LP problems that would be solved by selecting $n_2$ greater than 1 at depths d = 1 and 2, for example, which would add the same constant number to both totals. Likewise, we are disregarding the effort of the Winnowing Procedure for isolating the $n_2$ candidate variables, which would also add a constant (presumably somewhat smaller) in both cases.

If at any point in applying (2a) the nodes retained are all descendants of just one of the two nodes $(LP_h^+)$ and $(LP_h^-)$ generated from the root node $(LP^o)$, then the process can stop because this automatically singles out the initial up or down branch that will be selected at $(LP^o)$.

We may also illustrate the outcome of employing the Best sibling option of (2b) using the same values of $d_o = 2$ and Lim(d) = Lim = 3. In this case, the steps up through the stage of selecting 3 pairs of sibling nodes from the 4 generated in going from d = 2 to d = 3 are the same as before. However, by option (2b) we only retain the best node (the one yielding $Min(Eval_j^+, Eval_j^-)$) from each of the 3 pairs at depth 3, and then these 3 nodes generate 3 more pairs in going from d = 3 to d = 4, and from these 6 nodes we again select 3 to produce a "steady state" where the number of nodes generated and examined for D = 6 is given by 2 + 4 + 6 + 6 + 6 + 6 = 30.

The use of (2b) causes each of the continuations of the tree after $d = d_o$ to consist of a single path, without further branching.

Finally, (2c) creates the same kind of single path continuation as (2b), except that we may select both sibling nodes or 0 sibling nodes from the sibling node pairs produced at each depth d + 1, subject to selecting Lim(d) nodes in total.

The reduction in the number of nodes generated and examined for (2a), (2b) and (2c) suggests the option of exploring the narrow gauge tree to greater depths, or alternatively, of generating more than one such tree. Additional simple ways to reduce computation as a companion to these approaches are given in Appendix 1.

### 4.4 A Simplified Narrow Gauge Strategy for D = 2

If the look-ahead tree is reduced to a total depth of D = 2, we may apply a strategy that is roughly the opposite of the post-winnowing approach that effectively performs very little winnowing at all. More precisely, in this simplified case we cut back greatly on the work done in Stages 1 and 2 of the Winnowing Procedure (as by selecting the number of pivots $k_2$ to be very small, or even by accepting all $n_1$ variables passed along from Stage 1). Let $n_2(d)$ denote the value of $n_2$ selected at the depths d = 0 and d = 1 as a basis for creating the tree whose leaf nodes are at D = 2. The total number of LP problems we will solve to optimality is therefore $2n_2(0)$ and $4n_2(1)$ in order to select the branching variables at depth d = 0 and d = 1. Since the Winnowing Procedure is greatly reduced, almost all effort will result from solving these LP problems, and we may ration this effort to make it similar to the effort of solving 2·|F| LP problems by strong branching if we choose $n_2(0) + 2n_2(1)$ to be approximately |F|. Natural candidate values would be $n_2(0) = v \cdot n_2(1)$ for v ranging between 1 and 2. (For v = 1, $n_2(0) = |F|/3$ and for v = 2, $n_2(0) = |F|/2$.)

Such a D = 2 approach could then use the ideas underlying Criterion 7 of Section 2 where the unit costs for a node at D = 2 come directly from the parent node at d = 1, but using unit costs from basic variables at d = 0 (where possible) to avoid using reduced costs from non-basic variables at d = 1.

The goal of establishing $n_2(0)$ and $n_2(1)$ for this simplified D = 2 approach would not be to match the amount of computation produced by strong branching, however, since the benefits of looking ahead even for such a small depth may make it possible to attain the objective of finding an optimal solution somewhat earlier by permitting the computation to be modestly greater. Once more, we place greater emphasis on finding an optimal (or near optimal) solution more rapidly, in contrast to verifying the optimality of this solution by a tree search that exhausts all relevant alternatives. From this orientation, we can likewise employ a special D = 3 option that chooses a value $n_2(2)$ that is larger than would normally be considered, following a similar pattern to that described above. To prevent the approach from consuming an excessive amount of computation, it would be limited to being applied only a restricted number of times. Such an approach is relevant to exploiting the notion of persistent attractiveness, as discussed in Section 7.1.

## 5. Branching on Derivative Variables and Branch Reversals

We examine two main types of additional strategies, the first taking advantage of deeper branching by means of branching on derivative variables and the second identifying possibilities for reversing branches previously made. By "deeper branching" we refer to penetrating deeper into the solution space at each level of the narrow gauge tree, rather than to increasing the value of D.

### 5.1 Branching on Derivative Variables: Exploiting Straddle Branching

A strategy of branching on disjunctions other than those that compel a fractional *x* variable to be rounded up or down has been proposed in various forms over the years. However, none of these disjunctive branching approaches has been considered preferable in commercial implementations to the simple procedure of branching on $x_j \geq \lceil x_j^o \rceil$ and $x_j \leq \lfloor x_j^o \rfloor$ (as by the device of setting $L_j^+ = \lceil x_j^o \rceil$, and $U_j^- = \lfloor x_j^o \rfloor$ in the child problems (LP$^+$) and (LP$^-$)).

We likewise endorse the simple branching disjunction given by $x_j \geq \lceil x_j^o \rceil$ and $x_j \leq \lfloor x_j^o \rfloor$, but suggest the use of branching on a different disjunction for the purpose of (1) generating stronger bounds and (2) identifying the particular variable $x_k$ that should advantageously be selected for branching. A disjunction that accomplishes these ends is the *straddle branching* approach which executes a straightforward integer transformation on $x_j$ to yield a variable $z_j$ such that both of the inequalities $z_j \geq \lceil x_j^o \rceil$ and $z_j \leq \lfloor x_j^o \rfloor$ dominate the corresponding simple inequalities $x_j \geq \lceil x_j^o \rceil$ and $x_j \leq \lfloor x_j^o \rfloor$ (Glover and Laguna, 1997).

Related to, but different than the transformation that creates the mixed integer cut of Gomory (1963),[8] the transformation yielding the branching variable $z_j$ produces a disjunction such that each member of $z_j \geq \lceil x_j^o \rceil$ and $z_j \leq \lfloor x_j^o \rfloor$ cuts off a portion of the feasible LP region that strictly includes the space eliminated by the Gomory cut. It can be shown geometrically that the branching inequalities for $z_j$ "straddle" the Gomory cut relative to its intersections with the nonnegative orthant of the nonbasic variables, hence giving rise to the straddle branching name. The fact that $z_j$ is derived from $x_j$ is an important feature, since branching on $z_j$ thus gives an indication of the relevance of branching on $x_j$. Consequently, the stronger branching relative to $z_j$ can be used as a mechanism for uncovering the broader impact of the $x_j$ branching.

We now derive the straddle branching inequalities as follows. Let NB($x$) denote the current nonbasic $x$ variables and NB($y$) denote the current nonbasic $y$ variables in the optimal solution to (LP$^o$) (or more generally in any LP extreme point solution to (LP$^o$)). Then, we consider a fractional basic variable $x_j$ in this solution, whose basic solution representation may be written as

$$x_j + \sum(a_{ji} x_i : i \in NB(x)) + \sum(d_{ji} y_i : i \in NB(y)) = x_j^o$$

(The $a_{ji}$ and $d_{ji}$ coefficients here of course do not correspond to entries of the $A$ and $D$ matrices in the definition of (MIP).) Now, create a new integer variable $z_j$ by reference to integers $q_i$, whose values will be determined below, and defining

$$z_j = x_j + \sum(q_i x_i : i \in NB(x))$$

Hence we may write

$$z_j + (a_{ji} - q_i) x_i : i \in NB(x)) + \sum(d_{ji} y_i : i \in NB(y)) = x_j^o$$

Since $z_j$ is an integer variable, it gives rise to the disjunction $z_j \geq \lceil x_j^o \rceil$ or $z_j \leq \lfloor x_j^o \rfloor$. The question becomes: how can this disjunction be made as strong as possible by choosing the $q_i$ coefficients properly?

---

[8] A $z_j$ branch cannot be obtained by generating a Gomory mixed integer cut from the corresponding $x_j$ branch. In addition, the slack variable for the Gomory cut is not an integer variable, so no disjunction can be derived from it (as in an attempt to first introduce a Gomory cut and then to branch on it). The $z_j$ branches on the other hand have the additional novelty of yielding integer-valued slack variables, hence allowing these variables to become incorporated into further disjunctions.

The answer, as shown in Glover and Laguna (1997), is as follows. Let $r_i$ and $s_i$ denote the fractional parts of the coefficients $a_{ji}$ given by $r_i = a_{ji} - \lfloor a_{ji} \rfloor$ and $s_i = \lceil a_{ji} \rceil - a_{ji}$. (Hence $r_i = s_i = 0$ if $a_{ji}$ is an integer and $r_i = 1 - s_i$ otherwise. Similarly, let $r_o$ and $s_o$ denote the corresponding fractional parts of $x_j^o$ (hence expressed in terms of our previous notation, $r_o = f_o^-$ and $s_o = f_o^+$). Then we partition $NB(x)$ into the sets:

$NB1(x) = \{i \in NB(x): x_i = L_i^o \text{ and } r_i \leq r_o \text{ or } x_i = U_i^o \text{ and } s_i \geq s_o\}$
$NB2(x) = \{i \in NB(x): x_i = L_i^o \text{ and } r_i \geq r_o \text{ or } x_i = U_i^o \text{ and } s_i \leq s_o\}$

Nonbasic variables with $x_i = L_i^o$ and $r_i = r_o$, or with $x_i = U_i^o$ and $s_i = s_o$, can arbitrarily be assigned to either of $NB1(x)$ or $NB2(x)$. Variables with $s_i = r_i = 0$ can be assigned to either of these sets without making a difference since they will vanish from the final branching representation.

Then, to create the desired straddle branches, we determine $z_j = x_j + \sum(q_i x_i: i \in NB(x))$ by defining $q_i = -\lfloor a_{ji} \rfloor$ for $i \in NB1(x)$ and $q_i = -\lceil a_{ji} \rceil$ for $i \in NB2(x)$. From this we obtain

$$z_j + \sum(r_i x_i: i \in NB1(x)) - \sum(s_i x_i: i \in NB2(x)) + \sum(d_{ji} y_i: i \in NB(y)) = x_j^o$$

Let $z^+$ and $z^-$ denote nonnegative integer slack variables for the up and down straddle branching inequalities $z_j \geq \lceil x_j^o \rceil$ and $z_j \leq \lfloor x_j^o \rfloor$, thus giving $z_j - z^+ = \lceil x_j^o \rceil$ and $z_j + z^- = \lfloor x_j^o \rfloor$. Solving for $z_j$ and substituting the result in the $z_j$ equation above yields the two straddle branching (S-B) equations to add (alternately) to the problem (LP$^o$):[9]

$z^+ + \sum(r_j x_j: j \in NB1(x)) - \sum(s_j x_j: j \in NB2(x)) + \sum(d_{kj} y_j: j \in NB(y)) = -s_o$ (S-B: Up branch)

$z^- - \sum(r_j x_j: j \in NB1(x)) + \sum(s_j x_j: j \in NB2(x)) - \sum(d_{kj} y_j: j \in NB(y)) = -r_o$ (S-B: Down branch)

Each of the nonnegative integer-valued slack variables $z^+$ and $z^-$ takes the role of a basic variable in its respective equation.

Due to the greater strength of these (S-B) branches in comparison to the $x_j$ branches, we propose using them in the narrow gauge strategy to determine implications of branching on $x_j$ (combined with the implications of the integer requirements for the nonbasic $x_i$ variables). Specifically, by this approach straddle branching is used in the Winnowing Procedure in Step (1) of the tree generation (replacing the exploratory branching on variables $x_j$ by straddle branching on associated variables $z_j$), and in the branching in Step (2) of the tree generation (replacing the branching on $x_h$ by straddle branching on the associated variable $z_h$). Finally, when the tree generation strategy selects a variable $x_k$ as the winning variable for branching in (LP$^o$), then this branch is executed in the customary way without recourse to straddle branching.[10]

---

[9] We have slightly extended the exposition of Glover and Laguna (1997), which does not include reference to nonbasic variables at their upper bounds.

### 5.2 Branch Reversals and Resistance Measures

During a sequence of dual pivots to solve the LP problem associated with a given branch, the infeasibility termination condition $x_o \geq x_o^*$ is often supplemented by fixing a nonbasic variable $x_j$ at its current lower or upper bound if its reduced cost $RC_j$ satisfies $|RC_j| + x_o \geq x_o^*$, since this implies $x_j$ cannot be changed by a full unit from its assigned bound. (Such a change adds a cost of at least $RC_j$ to $x_o$ when $x_j$ is at its lower bound and a cost of at least $- RC_j$ to $x_o$ when $x_j$ is at its upper bound.)[11] The fixing restriction, as indicated earlier for all compulsory restrictions, is attached to the node produced by the branch and is inherited by all descendants of that node.

However, as an alternative when generating the narrow gauge look-ahead tree, we disregard such opportunities to fix nonbasic variables at their bounds. Instead, we interpret reduced costs in a different way as a *measure of resistance* to the bound imposed, or in other words, as an indication of the attractiveness of reversing the branch that created this bound (Glover, 2004). Thus at the node for problem (LP$^o$), the reduced cost $RC_j$ for a nonbasic variable $x_j$ at its lower bound $L_j^o$ signals that the value $x_o^o$ has the potential to improve by (at most) the value $RC_j$ if the lower bound is removed (recovering the previous lower bound) and the upper bound is changed to $U_j^o := L_j^o - 1$. Correspondingly, a nonbasic variable $x_j$ at its upper bound $U_j^o$ signals that the value $x_o^o$ has the potential to improve by (at most) the value $- RC_j$ if the upper bound is removed (recovering the previous upper bound) and the lower bound is changed to $L_j^o := U_j^o + 1$.

A more ambitious evaluation of branch reversals can be carried out by investigating their consequences by means of linear programming. This can be initiated for the case where the upper bound is changed to $U_j^o := L_j^o - 1$ by adding the current updated LP nonbasic column for $x_j$ to the column of constants in the LP tableau, and for the case where lower bound is changed to $L_j^o := U_j^o + 1$ by subtracting the updated LP nonbasic column for $x_j$ from the column of constants, followed in both cases by replacing $x_j$'s column by its negative. After initiation, the dual simplex method can be applied to complete the LP update.

We consider the option of executing a branch reversal at a leaf node of the look-ahead tree (at depth D) by selecting a variable $x_j$ that has the largest $|RC_j|$ value at this node, subject to requiring $|RC_j| \geq T$ for a chosen threshold T. Such a threshold can be based, for example, on a convex combination of the average and maximum $|RC_j|$ values over all leaf nodes.

---

[10] One merit of branching on $x_k$ rather than $z_k$ is the ability to avoid introducing an additional nonnegative integer variable $z^+$ or $z^-$ at each branch. Such variables need not be accumulated since they can be dropped as soon as they become nonbasic, which may lose the benefit of some of their influence but does not affect the validity of solutions generated. Since the narrow gauge tree will only accumulate a few such variables, the issue of dropping them is unimportant in this case.

[11] An implementation that avoids checking whether nonbasic variables are at their lower or upper bounds is to update variables by translation and complementation operations so that their current instances are always nonnegative and at a lower bound of zero when nonbasic. In this case, current reduced costs during dual LP iterations are always nonnegative.

This branch reversal strategy must be restricted to branches that lie in the look-ahead tree rather than including branches of B&B tree itself, unless the branching process for the B&B tree likewise avoids imposing implied bounds. More generally, a branch can be reversed only by dropping all implied restrictions accumulated on the path from the branch's child node to the node where the reversal occurs. Within the setting of the full B&B tree, such reversals provide an opportunity to overcome the rigid adherence to the tree structure that sometimes can lock an MIP B&B method into an unproductive search, particularly given the fact that earlier decisions are made on the basis of less information than later ones. A fuller discussion of branch reversals occurs in the context of tabu branching in Glover and Laguna (1997).

## 6. Analytical Branching – New Forms and Uses of Pseudo-costs.

The narrow gauge strategy can be implemented by making use of pseudo-costs to replace evaluations produced by solving LP problems, in the same manner as done in common variants of strong branching and reliability branching approaches. However, a consideration of the motives and circumstances under which pseudo-costs are employed also invites the creation of new forms and uses of these costs. To set the stage for this, we first briefly sketch how pseudo-costs are commonly constructed and used in current branching strategies.

### 6.1 Pseudo-cost Background

As before, we let $(LP_j^+)$ and $(LP_j^-)$ respectively denote the two LP problems derived from $(LP^o)$ by the up and down branches that set $L_j^+ = \lceil x_j^o \rceil$ and $U_j^- = \lfloor x_j^o \rfloor$ for a fractional variable $x_j$. Likewise, letting $x_{oj}^+$ and $x_{oj}^-$ denote the optimum $x_o$ values to these latter problems, we define unit costs $UC_j^+$ and $UC_j^-$ as in Section 3:

$$UC_j^+ = (x_{oj}^+ - x_o^o)/f_j^+ \text{ and } UC_j^- = (x_{oj}^- - x_o^o)/f_j^-.$$

Then the pseudo-costs for variable $x_j$, which we denote by $PseudoC_j^+$ and $PseudoC_j^-$ are the historical averages of such unit costs by reference to the number of problems $n_j^+$ and $n_j^-$ in which $x_j$ was chosen as the branching variable and in which the solution to $(LP_j^+)$ and $(LP_j^-)$ (respectively) was LP feasible. Thus, using summations over the problems giving rise to the counts $n_j^+$ and $n_j^-$:

$$PseudoC_j^+ = \sum(UC_j^+)/n_j^+ \text{ and } PseudoC_j^- = \sum(UC_j^-)/n_j^-$$

If $n_j^+$ or $n_j^-$ is 0, the associated value $PseudoC_j^+$ or $PseudoC_j^-$ is defined to be 0.

Based on this, a pseudo-cost score function is created to estimate the LP solution cost of a current up or down branch, hence to provide an estimate of the evaluations $Eval_j^+$ and $Eval_j^-$ that would result by solving the problems $(LP_j^+)$ and $(LP_j^-)$. Referring to this score function as $PseudoEval_j^+$ and $PseudoEval_j^-$, we obtain

$$PseudoEval_j^+ = PseudoC_j^+ \cdot f_j^+ \text{ and } PseudoEval_j^- = PseudoC_j^- \cdot f_j^-$$

The values $f_j^+$ and $f_j^-$ in this formula refer to the current problem $(LP^o)$ while those in the formula for $UC_j^+$ and $UC_j^-$ refer to past LP problems where $x_j$ was chosen as the branching variable and where the associated up or down branch produced a feasible LP solution.

### 6.2 Proxy Costs and Analytical Pseudo-Costs

Analytical Branching is based on *analytical pseudo-costs* which are produced in a different manner than customary pseudo-costs. These pseudo-costs can be used both with narrow gauge strategies or independent of such strategies.

We first present the basic ideas underlying analytical branching, and then examine ways in which these ideas can be implemented.

Let *proxy-cost* refer an estimate of the $x_o$ value, $x_o^+$ or $x_o^-$, for the child node $(LP_j^+)$ or $(LP_j^-)$ of the current node $(LP^o)$. Ostensibly, the goal of the pseudo-costs generated in the pseudo-cost literature is to identify good proxy-costs, that is, which are as accurate as possible in identifying $x_o^+$ and $x_o^-$, or more particularly the incremental values $x_o^+ - x_o^o$ and $x_o^- - x_o^o$.

As a basis for identifying better proxy costs, we make reference to the *Extended B&B tree* which consists of all branches explored by solving LP problems throughout the history of the B&B process up to the present point, including branches evaluated but not taken. For clearer differentiation, we refer to the B&B tree itself (consisting only of branches taken) as the *Basic B&B tree*. (Consequently, the Extended B&B tree, unlike the Basic B&B tree, is not a binary tree with two branches leaving each node.)

Within the Extended B&B tree, we define the branches on different variables $x_j$ all of which derive from the same node $(LP^o)$ as *sibling branches*. (Thus each sibling branch gives rise to its own pair of sibling nodes, that differs from the pair of sibling nodes at all other sibling branches, but whose associated branches are interrelated by virtue of issuing from a common parent.) Finally, we refer to those branches of the Extended B&B tree that do not lie in the Basic B&B tree as *tentative branches* and refer to their associated leaf nodes as *tentative nodes*.

To give a preliminary indication of the direction we wish to take in creating analytical pseudo-cost, we note that the idea behind Criterion 7 in Section 2 is motivated by the intimate relationship between sibling branches. Specifically, we conjecture that the unit cost values $UC_j^+$ and $UC_j^-$ computed for tentative sibling nodes $(LP_j^+)$ and $(LP_j^-)$ of $(LP^o)$ – by themselves, without reference to unit cost values computed at other nodes – provide useful cost estimates for these same $x_j$ branches after first branching on the particular variable $x_k$ which is chosen to be the branching variable at the node ($LP^o$).

The logic is that $x_k$ and the various $x_j$ variables of its sibling branches all share the same inheritance (the same collection of predecessor branches) in the Basic B&B tree up until the point where $x_k$ is chosen for branching, and hence the major part of the influence of prior decisions in creating the Basic B&B tree is the same for all of them. This shared influence suggests that pseudo-costs created by the rule underlying the definitions of $MinCost_j^+$ and $MinCost_j^-$ (embodied in Criterion 7), should be more relevant than by referring to other nodes – produced by more distant up and down branches in the Extended B&B tree – as in customary pseudo-cost calculations based on averages (as described in Section 5.1).

To make this idea more precise and more general, we introduce the following terminology.

The *processor path*, denoted Path(u), relative to a node u of the Extended B&B tree, is the path of nodes and edges from node u to the root node of the Extended B&B tree (which is the same as the root node for the Basic B&B tree). The number of edges on Path(u) is denoted by |Path(u)|.

For two nodes u and v of the Extended B&B tree, define Intersect(u,v) = Path(u) ∩ Path(v). Note that Intersect(u,v) lies in the Basic B&B tree and is a predecessor path Path(r) where r is the first node of their intersection (farthest from the root). Hence |Intersect(u,v)| denotes the number of edges on this path (which may be 0 in case r corresponds to the root node).

Dif(u,v) = Path(u)\Path(v) (hence Path(u)\Intersect(u,v)). By convention, we allow Dif(u,v) to include the node r for which Path(r) = Intersect(u,v).

Finally, let SymDif(u,v) = Dif(u,v) ∪ Dif(v,u) (the symmetric difference of Path(u) and Path(v)), which is a path in the Extended B&B tree between nodes u and v that excludes the intersection of these paths.

The motive for this terminology is based on supposing that u and v denote nodes produced by branching in the same direction on the same variable $x_j$ (and optionally, for greater refinement, by imposing the same lower or upper bound). We suppose that node u was generated earlier in the history of generating the Extended B&B tree, and node v is a node we currently wish to evaluate, before solving its associated linear program. (Hence, strictly speaking, node v at present lies outside the Extended B&B tree, although its parent v' lies in the Basic B&B tree. On the other hand, node u lies in the Extended B&B tree and may or may not lie in the Basic B&B tree.) We want to estimate the objective function value $x_o(v)$ at node v (or more precisely, the value $x_o(v) - x_o(v')$) that would result by solving its associated linear program, and wish to determine the relevance of the unit cost value $UC_j$ (= $UC_j^+$ or $UC_j^-$) at node u for this purpose.

The following three hypotheses are the foundation for creating analytical pseudo-costs.

*Hypothesis 1*: The accuracy of using $UC_j$ from node u to estimate $x_o(v)$ increases with the value of |Intersect(u,v)| and decreases with the value of |SymDif(u,v)|.

The interpretation of Hypothesis 1 is based on viewing Intersect(u,v) as composing the branches of their "shared inheritance" – i.e., the branches which constitute constraints shared by the problems LP(u) and LP(v) at nodes u and v – and viewing SymDif(u,v) as composing the branches (and hence constraints) by which this inheritance differs. Consequently, if node s is another node of the Extended B&B tree that likewise results by branching in the same direction on the same variable $x_j$, then node v dominates node s by the measure of Hypothesis 1 if |Intersect(u,v)| ≥ |Intersect(s,v)| and |SymDif(u,v)| ≤ |SymDif(s,v)|. For a simple approximation, we may consider that the accuracy of using $UC_j$ from node u is likely to be greater than the accuracy of using $UC_j$ from node v if |Intersect(u,v)|/|SymDif(u,v)| ≥ |Intersect(s,v)|/|SymDif(s,v)|. It is important to include compulsory branches in the definitions of Path(u) and Path(v), because these can potentially increase the size of Intersect(u,v) or SymDif(u,v).

We define analytical pseudo-costs to be those computed by restricting attention to non-dominated nodes u as identified the measure of Hypothesis 1. Then Hypothesis 1 leads directly to our second hypothesis.

*Hypothesis 2*: Pseudo-costs computed in the customary manner, based on averages involving different unit costs, will not provide estimates for $x_o(v)$ as accurate as those based on analytical pseudo-costs.

Hypotheses 1 and 2 suggest the creation of a maximum acceptable threshold value MaxSymDif and a minimum acceptable threshold value MinIntersect, to be determined by experimentation, according to the following hypothesis.

*Hypothesis 3*: There exist threshold values MaxSymDif and MinIntersect which will assure an appropriate degree of accuracy in evaluating $x_o(v)$ using the unit cost $UC_j$ at node u if |SymDif(u,v)| ≤ MaxSymDif and |Intersect(u,v)| ≥ MinIntersect.

Stated differently, a tentative branch should be evaluated by solving a linear program rather than relying on unit cost values when there is no node u that permits the threshold conditions of Hypothesis 3 to be satisfied. This implies that the usual policy of not employing pseudo-costs close to the root of the B&B tree is a proper one, but also implies that unit costs should not be generated close to the root rather than incorporating such costs into averages to compute later pseudo-costs. (By extension of Hypothesis 3, we may also refer to a value MinRatio and require |Intersect(u,v)|/|SymDif(u,v)| ≥ MinRatio as a condition for deciding the appropriateness of using a unit cost $UC_j$ as in Hypothesis 3.)

In general, the foregoing hypotheses give a means for selecting preferable $UC_j$ costs as a basis for computing pseudo-costs, and also suggest a way to determine when no such unit costs qualify to be used, and an LP problem should be solved instead.

As a further connection to Criterion 7 of Section 2, we observe that the unit cost values used in this criterion are implicitly based on $SymDif(u,v) = 3$, which is the smallest possible value SymDif can achieve. (Specifically, if $x_j$ is the source of a tentative branch leading to node u, and $x_k$ is the selected branching variable from their common parent node, and we wish to evaluate an $x_j$ branch to node v after branching on $x_k$, then node u and node v must be separated by 3 branches. Likewise, it is easy to see this scenario maximizes the value of Intersect(u,v) relative to evaluating the indicated node v.)

### 6.3 Rules for Identifying Best Nodes for Creating Analytical Pseudo-Costs

The relationships involving $|Intersection(u,v)|$ and $|SymDif(u,v)|$ underlying analytical pseudo-costs are more readily tracked in B&B trees generated by depth first branching than by other branching protocols.

In depth first branching, after a backtracking operation, only the linear programming $UC_j$ evaluations made at the node closest to the node q where a complementary branch is taken (including node q in this determination) is relevant for consideration as node u of the foregoing hypotheses.

During forward steps that grow the B&B tree without backtracking, the only node u that is relevant to consider is the one most recently examined to compute $UC_j$ by linear programming. (Different variables $x_j$ may of course have their $UC_j$ values computed at different nodes.)

Each time a new $UC_j$ determination is made at some node u during a sequence of forward steps, the value $|Intersect(u,v)|$ increases and the value $|SymDif(u,v)$ decreases, relative to any node v beyond u in this sequence, hence by Hypothesis 1 improving an analytical pseudo-cost determined by using this $UC_j$ value at node v. This suggests a strategy for saving memory when generating a depth first B&B tree that only keeps track of $UC_j$ values and the nodes where they are computed on the path from the root to the current leaf node $v_o$ of the B&B tree, since node v will be reached by a branch taken from $v_o$ and the relationship can easily be exploited where analytical pseudo-costs are improving for nodes u on the path $Path(v_o)$ that lie closer to the leaf node $v_o$ (and hence closer to v). Depending on when such a $UC_j$ update is made, it is possible that a dominating value may occur on a tributary branch of the B&B tree rather than on $Path(v_o)$, though such a possibility may be disregarded for expediency.

A final consideration to keep in mind is that it may be well to modify the threshold values of Hypothesis 2, or disregard them altogether, as the B&B process approaches more closely to a leaf node of the B&B tree where a feasible MIP solution is found, or where an infeasibilty is expected to occur. The reason is the same as enunciated earlier, where LP solution information about the impact of branching decisions becomes more accurate later in the tree (after more branches have been made), and hence it can be worth resorting to solving LP problems rather than using pseudo-costs at these late stages in order to wrap up the solution process more quickly. This creates the oddly asymmetrical approach of employing LP solutions more frequently at both early and late segments of the tree, compared to middle regions. Whether such a possibility translates into a useful strategy remains to be determined.

## 7. Global Relevance of Branching Decisions: Persistent Attractiveness and Reference Sets

The foregoing analysis of the relevance of branching decisions may be viewed as based on *local relevance*, which focuses on approximating the value $x_o(v) - x_o(v_o)$ that would be produced by solving a linear program at a current node $v_o$ which is the parent of a tentative node $v$.

We also may consider an evaluation based on *global relevance*, which refers to the merit of a branch that may ultimately lead to a high quality MIP solution, rather being driven evaluations implicitly or explicitly focus on the outcome of solving a local LP problem at a particular point in the tree. This global perspective requires a different type of memory than utilized in the construction of pseudo-costs, drawing on memory-based strategies of the type employed in tabu search. Two strategies in particular from this setting that invite adaptation to MIP branching decisions derive from methods for exploiting *persistent attractiveness* and *reference sets* composed of high quality solutions.

### 7.1 Persistent Attractiveness.

The Principle of Persistent Attractiveness (see e.g. Glover and Laguna, 1997) says that good choices derive from making decisions that have often appeared attractive, but that have not previously been made within a particular phase of search. That is, persistent attractiveness also carries with it the connotation of being "persistently unselected" within a specified domain or interval, and indicates the desirability of giving inducements to persistently attractive moves that upgrade the chance they will be selected.

The key idea is that a persistently attractive choice should be executed – whether or not it was ultimately selected – by moving it back to an earlier stage of the decision process when it first became attractive as one of the high quality moves available for consideration. In the case where the choice was ultimately selected, this re-positioned selection remains compatible with choosing the same collection of moves as before, and also offers an opportunity to make other moves. If the move was eventually selected anyway, there is nothing lost by selecting it earlier. But if it was not previously selected, the opportunity arises to make a more significant change in the search process.

The persistent attractiveness principle has an additional dimension within tree search, particularly when generating a look-ahead tree as in the narrow gauge branching strategy. Given that one of the two $x_k$ branches selected at any given node must be the "correct" branch (leading to the best possible solution that can be reached as a descendant of that node), if branching in a particular direction on another variable $x_j$ is attractive as a continuation of *both* $x_k$ branches, we may regard it as persistently attractive in an extended sense. If this branching option for $x_j$ was available at the same node where $x_j$ was selected, and even better, if the same up or down branching direction was preferred at that point, then a plausible strategy is to branch on $x_j$ instead of $x_k$. Such a strategy is directly available to the narrow gauge branching approach. For example, the special $D = 2$ and $D = 3$ options described in 4.4, are eminently suitable to applying this approach, where for $D = 3$ the method immediately intervenes upon reaching $d = 2$ if a persistently attractive variable $x_j$ has been found at this point.

With this in mind, we sketch a more general strategy for taking advantage of the persistent attractiveness notion when generating the narrow gauge tree.

We create a persistent attractiveness measure UpAttract(j) and DownAttract(j) for each fractional variable $x_j$ that counts the number of times $x_j$ qualifies both as attractive and preferable for branching up or down respectively (i.e., giving one of the highest evaluations at the node where $x_j$ is eligible to be a branching variable). These measures are set to 0 at the beginning of each iteration that launches the construction of the narrow gauge tree (hence at the beginning of the Basic Tree Generation Framework). Then they are incremented by 1 for the appropriate up or down direction for each of the $n_1$ branching variables $x_j$ in the set $F_1$ of the Winnowing Procedure, excluding the subset $F_2$ of $n_2$ variables whose branches are evaluated by solving the problems (LP$^+$) and (LP$^-$). For these latter variables, UpAttract(j) or DownAttract(j) is incremented by according to which branch is determined preferable by the LP solution.

Then, in a highly simplified approach to exploiting these measures, instead of branching on the original branching variable $x_k$ identified at depth $d = 0$ of the narrow gauge tree, we may branch on a variable $x_j$ from the set $F_2$ at $d = 0$ that yields the largest value AttractValue = Max(UpAttract(j), DownAttract(j)), selecting the up or down branching direction according to which of UpAttract(j) or DownAttract(j) is larger. This approach may be applied only if AttractValue exceeds a minimum threshold.

A somewhat more refined approach is to differentiate among the two branches for the initially selected variable $x_k$ at $d = 0$ by keeping a record UpAttract$^+$(j) and DownAttract$^+$(j) that applies to all evaluations performed the initial (LP$^+$) node (at $d = 1$) and its descendants, and a corresponding record UpAttract$^-$(j) and DownAttract$^-$(j) that applies the initial (LP$^-$) node and its descendants. All measures UpAttract$^+$(j), DownAttract$^+$(j), UpAttract$^-$(j) and DownAttract$^-$(j) include the counting increments at the node (LP$^o$) at $d = 0$. Thus, in particular, the measures UpAttract$^+$(j) and DownAttract$^+$(j) reflect the attractiveness counts for the "half-tree" determined by the initial node (LP$^o$) and its initial offspring (LP$^+$), and similarly UpAttract$^+$(j) and DownAttract$^+$(j) reflect the attractiveness counts for the half-tree determined by the initial node (LP$^o$) and its initial offspring (LP$^-$).

We differentiate between evaluations for the half-trees produced by the branches for the original $x_k$ variable because we are interested in the particular half-tree that contains the highest evaluation leaf node (or sibling leaf nodes) at depth D – i.e., the leaf node that determines the preferred branching direction for $x_k$ by the narrow gauge strategy as described in Section 4. In other words, persistent attractiveness in the alternative half-tree is considered less relevant. Restricting attention in this manner, we then select the branching variable $x_j$ instead of $x_k$ in the same manner as in the simplified approach above, where AttractValue now refers to counts in the indicated half-tree.

This type of approach can also be used to re-start a branch and bound process by keeping attractiveness measures produced just by nodes generated in the B&B tree, and then re-selecting the very first branching variable of the tree using these measures. Re-starting strategies are commonly used to overcome the disadvantage of being locked into the choice of the first (or first few) branching variables, and we hypothesize that the use of persistent attractiveness affords a way to employ such strategies more effectively. In fact, an initial extended form of the narrow gauge strategy, using the depth-based value $n_2(d)$ as described in the special D = 2 and D = 3 approaches of Section 4.4, but employing a one-time implementation for a larger D at the root of the B&B tree, allows a re-structuring of the tree without having to employ a more extensive B&B search before launching the re-starting process.

### 7.2 Reference Sets of High Quality Solutions

We anticipate the value of maintaining a reference set of high quality solutions as proposed in tabu search and scatter search, and which subsequently also has become an emphasis of so-called "elite population management" approaches in genetic algorithms. Our present use of reference sets is to create a bias toward selecting branches that would lead to solutions resembling those in the reference set, on the supposition that some of these branches are likely to be included in other high quality MIP solutions.

We stipulate that such a bias should be attenuated at greater depths of the B&B tree, where we expect global relevance should be superseded by local relevance. The reason for this supposition is similar to that underlying the disregard for the threshold values of Hypothesis 2 at greater depths of the tree. In particular, once a number of decisions have been made that favor branches leading to good solutions previously obtained, it becomes important to give greater attention to variations that lead to alternative solutions. The enumerative tree structure will automatically force solutions to differ from those obtained in the past, but in later stages of the tree it can be advantageous to allow greater leeway to local guidance to find solutions that are best.

We further stipulate that global relevance of a branch should increase if the branch leads to more than one high quality MIP solution. Hence we employ a modified evaluation that encourages the choice of a given branch based on the number of such solutions and a weighted average of their $x_o$ values.

To provide an evaluation measure that has an interpretation similar to that embodied in unit costs as employed in pseudo-cost evaluations, we create associated *global unit cost* values, denoted by $GUC_j$, derived from branches that lie on a path leading to a high quality MIP solution. However, we do not compute the $GUC_j$ values in the same manner as the $UC_j$ values, because then they would depend on the sequence of the branches involved, and a different sequence that led to the same solution would give different unit costs. Consequently, we calculate the $GUC_j$ values in a way that avoids this sequence-dependent outcome.

We begin by specifying a calculation for a single solution $x^r$ from a reference set R consisting of high quality (and MIP feasible) solutions. Let $x^o$ denote the LP solution at the root node of the B&B tree, and hence $\Delta_o = x_o^o - x_o^r$ identifies the total objective function change in moving from $x^o$ to $x^r$. Similarly, let $\Delta_j = x_j^r - x_j^o$ identify the change in the value of $x_j$ in moving from $x^o$ to $x^r$. Finally, let $n(r)$ be the number of variables for which $\Delta_j \neq 0$. Then the average $x_o$ change over these variables is $AvgCng = \Delta_o / n(r)$.

A naïve determination of the global unit costs can then be given by

If $\Delta_j > 0$: $GUC_j^+ = AvgCng/\Delta_j$, and $GUC_j^- = $ *Large*
If $\Delta_j < 0$: $GUC_j^- = AvgCng/|\Delta_j|$ and $GUC_j^+ = $ *Large*
If $\Delta_j = 0$: $GUC_j^+ = GUC_j^- = $ *Large*

Here *Large* is a special value whose meaning will be clarified subsequently. At present we note that a "large" value merely discourages taking a branch that does not lead to the high quality MIP solution $x^r$.

The foregoing calculation involves a potential distortion because branching on some variables will automatically create changes in other variables, and this may result in requiring fewer than $n(r)$ branches to go from $x^o$ to $x^r$. Consequently, when the solution $x^r$ is obtained, we perform a predecessor trace to identify the set of variables $N(r)$ that are chosen as branching variables on the path of the B&B tree from $x^o$ to $x^r$, excluding consideration of compulsory branches and also disregarding any branches such that $\Delta_j = 0$ (hence branches which were effectively cancelled as the result of other branches).

Hence, by this determination of $N(r)$ we determine the only "necessary" branches required for going from $x_o$ to $x^r$ as those that result in changing the variables $x_j$ for $j \in N(r)$ to their final values $x_j^r$. All remaining variables will then be compelled by the MIP inference rules to achieve their final $x_j^r$ values as well. (This is not completely accurate since the conditions under which inference rules are employed may vary within the B&B tree, but it will be approximately true if compulsory branches based on the requirement $x_o < x_o^*$ use an $x_o^*$ value at least as small as the one used when $x^r$ was generated.)

Then the foregoing naïve determination of global unit costs can be amended by defining $n(r) = |N(r)|$ and specifying

If $\Delta_j > 0$ and $j \in N(r)$: $GUC_j^+ = AvgCng/\Delta_j$, and $GUC_j^- = Large$
If $\Delta_j < 0$ and $j \in N(r)$: $GUC_j^- = AvgCng/|\Delta_j|$ and $GUC_j^+ = Large$
If $\Delta_j = 0$ or $j \notin N(r)$: $GUC_j^+ = GUC_j^- = Large$

**Computing $GUC_j$ values over all of the reference set R.**
We now consider the determination of $GUC_j$ values over all solutions $x^r$ in R. We'll first pursue a development that maintains a "unit cost orientation," and then show that it has certain shortcomings that lead to an alternative approach.

Applying the unit cost orientation, let $\Delta_j(r)$, $GUC_j^+(r)$ and $GUC_j^-(r)$ refer to the $\Delta_j$, $GUC_j^+$ and $GUC_j^-$ values for a given solution $x^r$, $r \in R$. For the following, it is useful to adopt the convention that $\Delta_j(r) = 0$ if $j \notin N(r)$.

Then we define composite $GUC_j^+$ and $GUC_j^-$ values by setting

$$GUC_j^+ = Min(GUC_j^+(r): r \in R) \text{ and } GUC_j^- = Min(GUC_j^-(r): r \in R)$$

We need to modify this based on the fact that we want $GUC_j^+$ ($GUC_j^-$) to be more attractive (hence smaller) as the number of solutions in which $\Delta_j(r) > 0$ ($\Delta_j(r) < 0$) increases.

Let $\delta_j^+(r) = 1$ for $\Delta_j(r) > 0$ and $\delta_j^+(r) = 0$ otherwise, and similarly let $\delta_j^-(r) = 1$ for $\Delta_j(r) < 0$ and $\delta_j^-(r) = 0$ otherwise. Then define

$$n_j^+ = \sum(\delta_j^+(r): r \in R) \text{ and } n_j^- = \sum(\delta_j^-(r): r \in R).$$

Hence, $n_j^+$ and $n_j^-$ identify the number of solutions $x^r$, $r \in R$, for which $\Delta_j(r) > 0$ and $\Delta_j(r) < 0$, respectively. Then, for a selected positive exponent, we replace the $GUC_j^+$ and $GUC_j^-$ values above by global cost values $GC_j^+$ and $GC_j^-$ given by

$$GC_j^+ = GUC_j^+/(n_j^+)^p \text{ and } GC_j^- = GUC_j^-/(n_j^-)^p$$

In this fashion the global cost decreases as a function of $n_j^+$ and $n_j^-$. We can now use $GC_j^+$ and $GC_j^-$ values in place of $UC_j^+$ and $UC_j^-$ values for computing pseudo-costs, with the understanding that these two ways of producing pseudo-costs are not commensurate. A discussion of potential limitations and further ramifications of such values appears in Appendix 2.

When using $GC_j^+$ and $GC_j^-$ values to guide the choice of branching variables and branches, we note that special monitoring is appropriate to determine the extent to which prior branching on a given variable $x_k$ has "eaten up" a reasonable allotment for this variable. (In the case of binary variables, this monitoring is unnecessary.)

Define the *effective branching distance* for the variable $x_j$, $j \in N(r)$, on the path from $x_o$ to the solution $x^r$ to be given by

$$BD_j(r) = |x_j^r - x_j^o|.$$

Let $R_j^+ = \{r \in R : \Delta_j(r) > 0\}$ and $R_j^- = \{r \in R : \Delta_j(r) < 0\}$
From this we can identify Min, Max and Mean effective branching distances for $x_j$ over the set $R_j^+$ by

$MinBD_j^+ = Min(BD_j(r): r \in R_j^+)$
$MaxBD_j^+ = Max(BD_j(r): r \in R_j^+)$
$MeanBD_j^+ = \sum(BD_j(r): r \in R_j^+)/|R_j^+|$

and similarly identify corresponding values over the set and $R_j^-$.

Then on any branching decision we only allow an up (down) branch for $x_j$ if the accumulated amount of branching in this direction does not exceed a selected convex combination of these values.

## 8. Conclusion

We have identified a collection of MIP branching strategies as a foundation for experimental study. These include new criteria for selecting branching variables, the use of Narrow Gauge Branching strategies based on generating look-ahead trees (accompanied by associated pre- and post-winnowing procedures), branching on derivative variables, strategies for reversing branches, the application of analytical branching based on relationships embodied in analytical pseudo-costs, and the incorporation of global relationships based on persistent attractiveness and reference sets.

Our proposals undertake to incorporate perspectives that have been useful in the design of effective metaheuristics and which we conjecture to be transferable to methods based on a branch and bound framework. The wide range of alternatives made available from such perspectives encourage future empirical research to determine the avenues that prove most fruitful.

### Acknowledgement:
We are indebted to Gregor Hendel for observations that have improved the exposition of this paper.We are indebted to Gregor Hendel for observations that have improved the exposition of this paper.

# References

Achterberg, T., T. Koch and A. Martin (2005) "Branching rules revisited," *Operations Research Letters*, Volume 33, Issue 1, pp. 42-54.

Applegate, D. L., R. E. Bixby, V. Chvatal and W. J. Cook (1995) "Finding cuts in the TSP (A Preliminary Report)," Technical RePort 95-05, DIMACS.

Benichou, M., J-M. Gauthier, P. Giroledet, G Hentges, G. Ribiere and O. Vincent (1971) "Experiments in mixed-integer programming," *Mathematical Programming*, Vol. 1, pp. 76-94.

Berthold, T. (2013) "Measuring the impact of primal heuristics," *Operations Research Letters,* Volume 41, Issue 6, pp. 611–614.

Berthold, T. and G. Hendel (2015) "Shift-and-Propagate," *Journal of Heuristics*, Vol. 21, pp. 73-106.

Blum, C. and A Roli (2003) "Metaheuristics in Combinatorial Optimization: Overview and Conceptual Comparison," *ACM Computing Surveys*, Volume 35 Issue 3, pp. 268-308.

Eckstein, J. and M. Nediak (2007) "Pivot, Cut and Dive: a heuristic for 0-1 mixed integer programming, "*Journal of Heuristics,* Volume 13, Issue 5, pp 471-503.

Crainic, T.G. and M. Toulouse (2003) "Parallel Strategies for Meta-Heuristics," Chapter 17 of *Handbook of Metaheuristics*, G. Kochenberger and F. Glover, eds., Kluwer Academic Publishers.

Eckstein, J. and M. Nediak (2007) "Pivot, Cut, and Dive: a heuristic for 0-1 mixed integer programming," *Journal of Heuristics*, Vol. 13, Issue 5, pp. 471-503.

Fischetti, M. and M. Monaci (2011) "Backdoor Branching," In O. Gunluck and G. J. Woeginger, eds, *Integer Programming and Combinatorial Optimization*, Vol. 6655 of *Lecture Notes in Computer Science*, pp. 183-191, Springer, Berlin.

Glankwamdee, W. and J. Linderoth (2011) "Lookahead Branching for Mixed Integer Programming," ICS 2011, 12th INFORMS Computing Society Conference Computing Society, _c 2011 INFORMS, isbn 978-0-9843378-1-1, doi 10.1287/ics.2011.0010, pages 130-147.

Glover, F. (1997) "A Template for Scatter Search and Path Relinking," in *Artificial Evolution, Lecture Notes in Computer Science,* 1363, J.-K. Hao, E. Lutton, E. Ronald , M. Schoenauer and D. Snyers, Eds. Springer, pp. 13-54.

Glover, F. (2006) "Parametric Tabu Search for Mixed Integer Programs," *Computers and Operations Research*, Volume 33, Issue 9, pp. 2449-2494.

Glover, F. and M. Laguna (1997) *Tabu Search*. Kluwer Academic Publishers.

Gomory, R. E. (1963) "An algorithm for integer solutions to linear programs," *Recent Advances in Mathematical l Programming*, R.L. Graves and P. Wolfe eds., McGraw-Hill, New York, pp. 269–302.

Hendel, G. (2015) "Enhancing MIP branching decisions by using the sample variance of pseudo-costs," *Integration of AI and OR Techniques in Constraint Programming: Lecture Notes in Computer Science Volume 9075*, pp. 199-214.

Li, X., O. Ergun and G. L. Nemhauser (2015) "A Dual Heuristic for Mixed Integer Programming," Preprint submitted to *Operations Research Letters*.

Linderoth, J.T. and M. W. P. Savelsberg (1999) "A computational study of search strategies for mixed integer programming," *Informs Journal on Computing*, Vol. 11, pp. 173-187.

Pryor, J and J. W. Chinneck (2011) "Faster integer-feasibility in mixed-integer linear programs by branching to force change," *Computers & Operations Research*, Vol. 38, No. 8, pp. 1143-1152.

## Appendix 1: Auxiliary Approaches to Reduce Computational Effort in the Narrow Gauge Method

We describe two simple auxiliary approaches to reduce computational effort in the narrow gauge method.

**Limiting the iterations for dual re-optimization**. A straightforward strategy to save computation is to impose a stronger limit on the maximum number of pivots allowed for solving the problems ($LP_h^+$) and ($LP_h^-$) in Step (2) of the Narrow Gauge Tree Generation process. This limit can be significantly smaller than one chosen to assure that the LP problems will almost always be solved to optimality. For example, the limit can be adaptively linked to stipulating that the greatest infeasibility (largest constraint violation) falls below a value $v_{lim}$, which can be calibrated to be some multiple $m < 1$ of the value $\text{Max}(\text{Min}(f_j^-, f_j^+))$ over the variables that are fractional on the current dual LP iteration.

We can additionally choose the limiting number of pivots to vary as a function of d. A decreasing function affords a greater reduction in computational expense by requiring fewer pivots at later levels of the tree. If the limit is applied only for nodes at depth $d = D - 1$ (to generate the final nodes at level D), there is no concern over selecting future branches out of the resulting leaf nodes and no need to refer to a value such as $v_{lim}$. In this case, since the number $2^D$ of nodes generated at level D is effectively half the number $2^1 + 2^2 + \ldots + 2^D = (2^D - 2) + 2^D$ of nodes for the complete tree, the limited number of pivots at this final stage will have a significant impact on the total computational effort.

**Using a Fixed Candidate List.** Computational effort can be additionally reduced by using a candidate list CList for branching variables that restricts consideration to a subset of those variables that were fractional at depth 0. In this approach, CList can be determined by reference to the Winnowing Procedure to consist of the $n_0$ variables chosen at Stage 0 for $d = 0$, hence constituting a subset of the fractional variables at the root node ($LP^o$). Then, each time the Winnowing Procedure is applied at Step (1) of the tree generation process, Stage 0 of the Winnowing Procedure for the current node under consideration only selects variables from CList. The tree generation terminates for the current node, making it a leaf node, if no variables in CList are fractional.

## Appendix 2: Potential Limitations and Ramifications of Global Unit Cost Evaluations.

Just as we have discussed potential limitations of traditional pseudo-costs and potential remedies based on the ideas of analytical branching in Section 6, it is useful to point out a limitation in the global unit cost evaluations introduced in Section 7.2.

To illustrate a difficulty that can arise in the definition of the $GUC_j^+$ and $GUC_j^-$ values, consider an exaggerated example consisting of two high quality solutions $x^1$ and $x^2$ where $x_o^1 \cong x_o^2$ and where $x^1$ results by changing the values of just two variables in going from $x^o$ to $x^1$, while $x^2$ results by changing the values of four variables in going from $x^o$ to $x^2$, hence $n(1) = 2$ and $n(2) = 4$. Since $x_o^1$ and $x_o^2$ are nearly the same, the value AvgCng for $x^1$ will be approximately twice that for $x^2$. Other things equal, the $GUC_j$ values for $x^1$ will then be roughly twice those for $x^2$, which means the values contributed by $x^2$ will appear twice as attractive as those derived from $x^1$. Yet, by contrast, the ability to reach a high quality solution by changing the values of only two variables versus changing the values of four variables would seem to warrant considering the $GUC_j$ values for $x^1$ to be more attractive than those for $x^2$, particularly for the purpose of using these values for selecting a variable to branch on. Changing the definition of the $GUC_j^+$ and $GUC_j^-$ values to refer to some form of average quantity does not repair this difficulty. On the other hand, the distortion disclosed in the foregoing example seems likely to be atypical, and may be less consequential than the type of distortion created in the use of ordinary pseudo-costs.

If a remedy to this potential limitation is sought, a conceivable solution would be to cancel out the effect of $n(r)$ by removing it from the definition of AvgCng (replacing AvgCng by $x_o^r - x_o^o$), but then the unit cost interpretation is broken. This may not be serious, since the replacement in Section 7.2 of the $GUC_j^+$ and $GUC_j^-$ values by the $GC_j^+$ and $GC_j^-$ values already severs the link to traditional pseudo-costs. However, we would still need a modification comparable to that involved in creating the $GC_j^+$ and $GC_j^-$ values.

Another option for going further is to make additional use of effective branching distances as introduced in Section 7.2 and define the (full) effective branching distance from $x^o$ to $x^r$ to be

$$BD(r) = \sum( BD_j(r): j \in N(r)).$$

Then $BD(r)$ might be incorporated into evaluations of branching decisions, under the supposition that the attractiveness of choosing $x_j$ as a branching variable increases for all $j \in N(r)$ as $BD(r)$ decreases. Such considerations for making branching decisions may properly be left to future exploration.